\documentclass[%
 amsmath,amssymb,
 aps,
 reprint
]{revtex4-1}

\usepackage{subfig}
\usepackage{siunitx}
\usepackage{graphicx}% Include figure files
\usepackage{dcolumn}% Align table columns on decimal point
\usepackage{bm}% bold math
\begin{document}

\preprint{APS/123-QED}

\title{Pattern Formation in Double-Layer Kerr Resonators with Coupled Modes}% Force line breaks with \\

\author{Antoine Bois$^*$}
%\email{antoine.bois@mail.utoronto.ca}
\author{Joyce K. S. Poon}%
\affiliation{%
The Edward S. Rogers Sr. Department of Electrical \& Computer Engineering, University of Toronto, Toronto, ON M5S 3G4, Canada \\
*Corresponding author: antoine.bois@mail.utoronto.ca
}%


\begin{abstract}

A double-layer Kerr resonator in which both coupled modes are excited and interact with each other via incoherent cross-phase modulation is investigated to reveal stable localized solutions beyond the usual formation mechanism involving a single mode. Periodic solutions from modulational instability are found to occur at a slight penalty on the nonlinear efficiency, but they stabilize the spatial dynamics, leading to dissipative solitons in previously unattainable regimes. Numerical simulations show paired breather solitons in addition to temporally stable solutions. The results demonstrate coupled modes can increase the stability of Kerr frequency comb generation.\end{abstract}

\maketitle

The study of localized light patterns in passive Kerr resonators has recently experienced a renewal of interest for its relevance to coherent frequency comb generation. In such resonators, dissipative Kerr solitons are able to arise spontaneously from a single monochromatic continuous-wave (cw) input. Their dynamics are well documented \cite{lugiato1987spatial, scroggie1994pattern,chembo2010modal, godey2014stability, parra2014dynamics,parra2016origin,hansson2016dynamics}. In short, in cavities with anomalous dispersion, dissipative solitons have been understood to emerge from the homoclinic orbit of a stable homogeneous steady state (HSS) passing asymptotically close, in phase space, to a stable periodic pattern, itself originating from modulational instability (MI). Compared to periodic patterns by themselves, solitons vastly broaden the spectrum of the optical field, making them preferred for frequency combs.

In a standard resonator, this soliton formation mechanism takes place with the lower intracavity power HSS in the bistable region of a nonlinear resonance. This metastability of the HSS causes stability issues in the practical implementations of Kerr combs \cite{yi2016active,karpov2016universal}. In the upper branch of the bistability, periodic patterns tend to progress into spatiotemporal chaos (sometimes also called optical turbulences) \cite{ikeda1980optical}, while in the low intracavity power monostable regime, spatial eigenvalues of the HSS are generally found unsuitable for this homoclinic orbit to occur, despite a HSS briefly co-existing with a subcritical periodic solution \cite{parra2014dynamics}.

In this Letter, we investigate the formation dynamics of both temporal dissipative solitions and periodic patterns in a resonator formed by two coupled guiding layers that support two waveguide modes of similar polarization. We show that incoherent cross-phase modulation (XPM) between the two modes leaves the onset of MI mostly intact and is able to modify the spatial eigenvalues to extend the range of input parameters that support heteroclinic coupling of a HSS to a periodic pattern. Our analysis accounts for both coupled modes simultaneously, in contrast to recent works that only study and use one of the coupled modes for dispersion engineering \cite{soltani2016enabling,kim2016frequency}.

\begin{figure}[!htb]
\centering
%\fbox{\includegraphics[width=\linewidth]{image4768}}
\includegraphics[width=\linewidth]{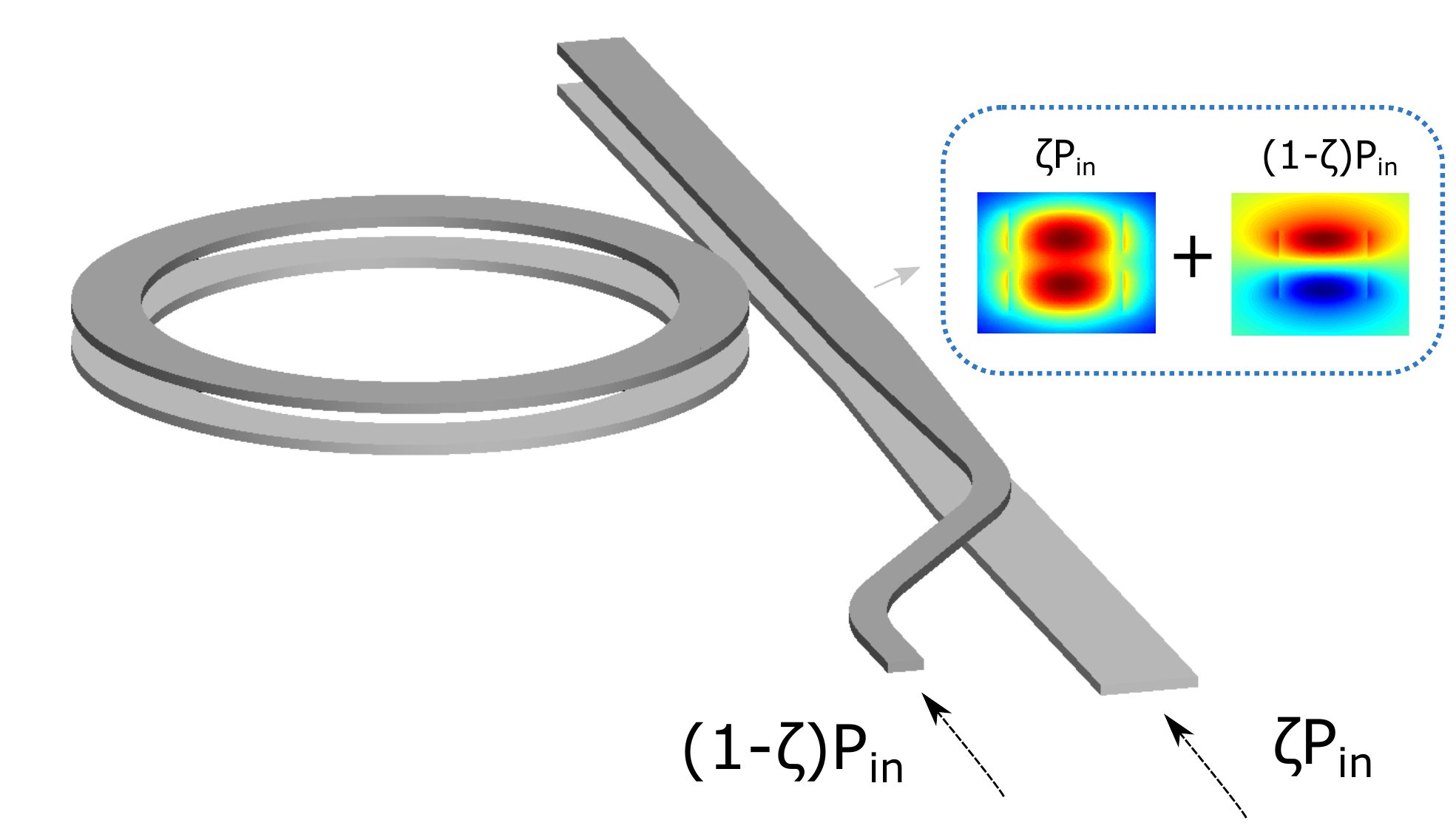}
\caption{A microring resonator consisting of two guiding layers supports symmetric and anti-symmetric modes for TE and transverse magnetic (TM) polarizations. Light launched into the wider (narrower) waveguide converts adiabatically to the  symmetric (anti-symmetric) mode of the vertical bus coupler for selective modal excitation. The waveguides then couple evanescently to the resonator with low cross-talk due to the phase matching imposed by the directional coupler. The splitting ratio, $\protect{\zeta}$, can be implemented either on- or off-chip. The optical input is monochromatic and has power $\protect{P_{in}}$.}
\label{fig:1}
\end{figure}

In our case, a coupled structure is employed mainly to circumvent the issue of modal anticrossings typical to multimode resonators. This phenomenon arises when two modes are simultaneously close to resonance, due to coupling between them that perturbs both their phases and amplitudes and thus prevents a resonance degeneracy, as described by standard coupled-mode theory \cite{liu2014investigation}. Here, we use the eigensolutions of the coupled propagation equations as a basis. In a double-layer resonator, this conveniently corresponds to the symmetric and anti-symmetric modes of the coupled structure. In addition to the disappearance of anticrossings, this basis offers an increased XPM efficiency compared to higher order modes, a relative ease of coupling from a bus waveguide, and a broad range of possible dispersion engineering enabled by the mode coupling term, $\kappa$, designed through the separation between and dimensions of the guiding layers \cite{soltani2016enabling,kim2016frequency}. Fig. \ref{fig:1} shows the proposed device in the form of a microring, including the input/output bus waveguide.  Such a device can be implemented in multilayer silicon nitride or silicon nitride-on-silicon integrated photonic platforms \cite{sacher2016multilayer,sacher2015multilayer}.%

Turning our attention to the dynamics of the system under study, we introduce the standard mean-field Lugiato-Lefever equation (LLE) \cite{lugiato1987spatial} to describe the nonlinear propagation of the slowly varying field envelopes. Using the normalization of  \cite{leo2010temporal}, modified to include the XPM interaction and mismatch in group velocities, the LLE becomes%
\begin{subequations}\label{eq:3}
\begin{equation}
\partial_t u = -(1 + i \theta_{u}) u + i\left(|u|^2 + 2|v|^2\right) u + S_{u} - i \eta_{u} \partial^2_\tau u - \delta \partial_\tau u,
\label{eq:3a}
\end{equation}
\begin{equation}
\partial_t v = -(1 + i \theta_{v}) v + i\left(|v|^2 + 2|u|^2\right) v + S_{v} - i \eta_{v} \partial^2_\tau v + \delta \partial_\tau v,
\label{eq:3b}
\end{equation}
\end{subequations}%
with $t$ as the slow time (on the scale of the optical round-trip time), $\tau$ as the fast time (in a reference frame traveling at the average of the group velocities), $\theta_{u,v}$ as the detuning with the closest linear resonance, $S_{u,v}$ as the amplitude of the coherent cw pump, and $\eta_{u,v}$ and $\delta$, the second-order dispersion and extraneous first-order dispersion respectively, being defined as a ratio to the average of the absolute values for the two modes. The first term on the right side relates to the normalized losses including propagation and coupling, here assumed similar for both modes. We choose to work in a strongly coupled regime such that the coherent four-wave mixing term $\exp\left(\pm i 4 \kappa t \right)$ oscillates rapidly with $t$ to average to zero and can be neglected. The mechanism presented here therefore depends only on \textit{incoherent} XPM.

\begin{figure}[!htb]
\centering
\captionsetup[subfloat]{justification=centering}
\subfloat[]{\includegraphics[width=0.5\linewidth]{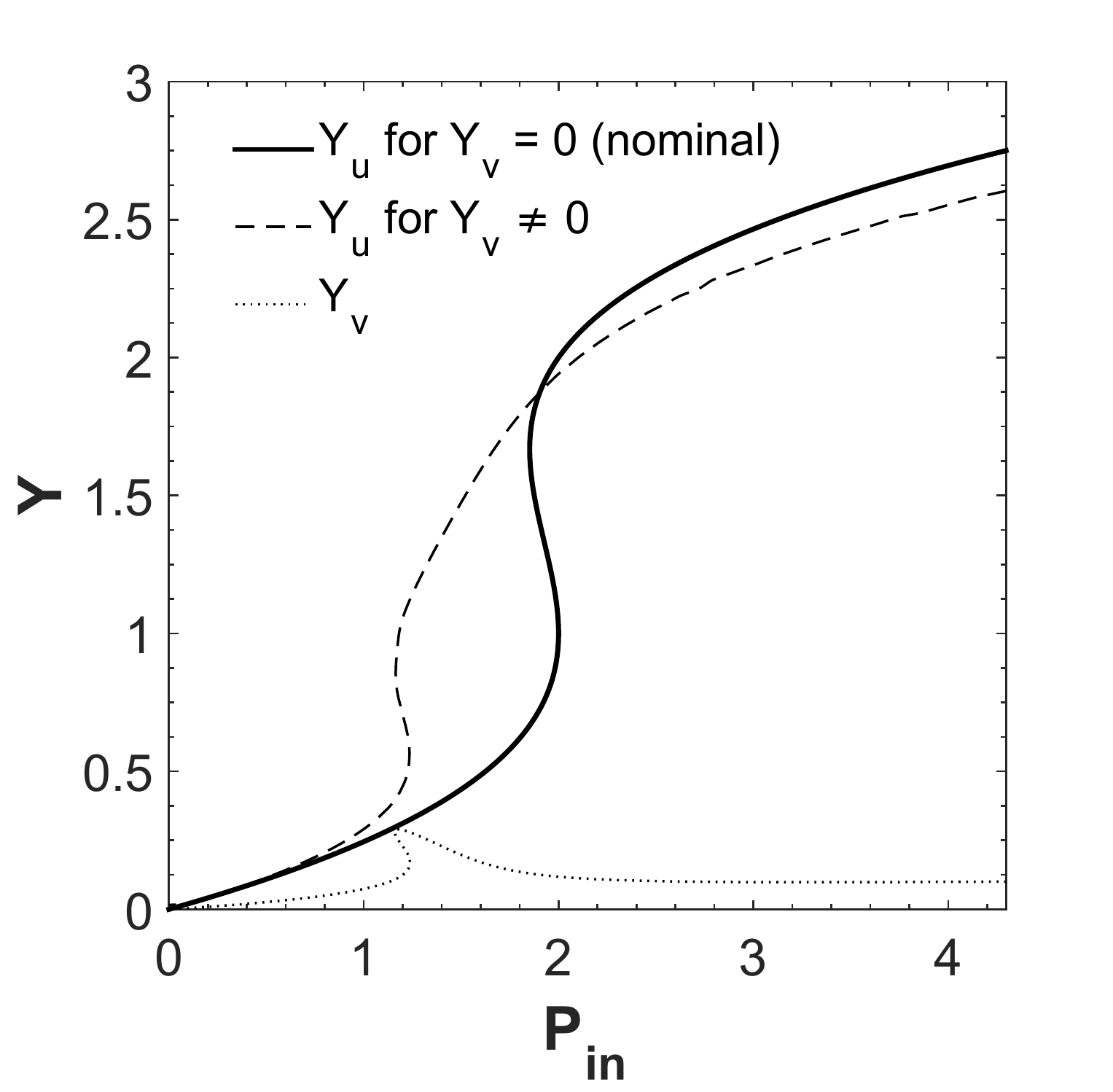}}
\subfloat[]{\includegraphics[width=0.5\linewidth]{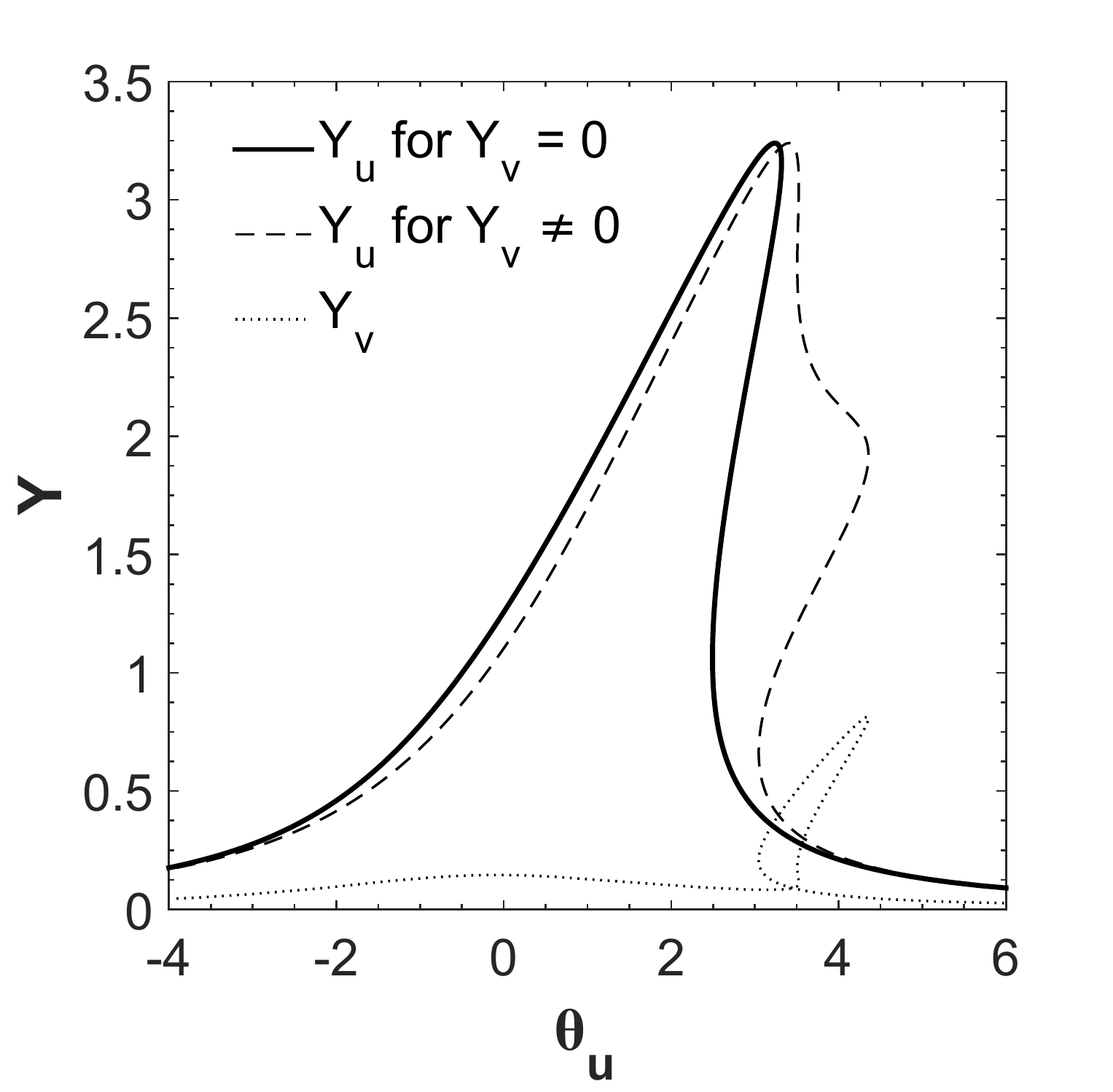}}
\caption{Intracavity power of the homogeneous solutions as a function of (a) input pump power $\protect{P_{in} = X_u + X_v}$, for $\protect{X_u / X_v = 3}$ and fixed detunings $\protect{\theta_u = 2}$, $\protect{\theta_v = \theta_u + 0.2}$; and (b) input detuning $\protect{\theta_u}$, for $\protect{\theta_v = \theta_u + 0.2}$ and $\protect{X_u = 3.24, X_u / X_v = 3}$.}
\label{fig:2}
\end{figure}

The homogeneous solutions of Eq. \ref{eq:3} are instructive as a starting point for the formation of patterned solutions. This is done by setting all derivatives to zero and leads to%
\begin{subequations}\label{eq:4}
\begin{equation}
X_{u} = Y^3_{u} - 2 \left(\theta_{u} - 2 Y_{v} \right) Y^2_{u} + \left[\left(\theta_{u} - 2 Y_{v}\right)^2 + 1\right]Y_{u},
\label{eq:4a}
\end{equation}
\begin{equation}
X_{v} = Y^3_{v} - 2 \left(\theta_{v} - 2 Y_{u} \right) Y^2_{v} + \left[\left(\theta_{v} - 2 Y_{u}\right)^2 + 1\right]Y_{v},
\label{eq:4b}
\end{equation}
\end{subequations}

\noindent where $X_{u,v} \equiv |S_{u,v}|^2$, the intensity of each cw input, and $Y_{u} \equiv |u|^2$, $Y_{v} \equiv |v|^2$. At first glance, the effect of XPM is to act as an additional effective detuning to $\theta_{u,v}$. However, the dynamical nature of $Y_{u}$ and $Y_{v}$ means that they are not independent of each other, and therefore neither is this control over the detuning. Approaching the resonance from the stable high frequency side ($\theta_{u,v} < 0$), despite $X_{u}$ and $X_{v}$ being potentially similar in magnitude, the mode with more power ends up overwhelmingly dominating the resonance. This is illustrated in Fig. \ref{fig:2}, where both are plotted on the same scale. %
This turns out to be a necessary condition for the stability of the HSS, at least to homogeneous perturbations. If both modes were to be sustained at close to full resonance, then any slight perturbation in any direction would strongly affect the XPM, in turn modifying the resonance condition and intracavity power of the second mode, and thus further destabilizing the original mode and amplifying that initial perturbation. This is formalized mathematically by writing small perturbations $\varepsilon \propto \exp\left(\lambda_t t + i \Omega \tau \right)$ into Eqs. \ref{eq:3a} and \ref{eq:3b}. Keeping only the linear terms and solving for the non-trivial solutions of $\left[\varepsilon_u, ~\varepsilon_u*, ~\varepsilon_v, ~\varepsilon_v* \right]^T$ leads to

\begin{widetext}
\small
\begin{equation}
\left|
\begin{aligned}
\begin{matrix}
-\lambda_t - \left(1 + i \theta_u \right) + i \left( \eta_u \Omega^2 - \delta \Omega \right) + 2i(|u|^2+|v|^2) & i |u|^2 & 2i |u| |v| & 2i |u| |v| \\ 
-i |u|^2 & -\lambda_t - \left(1 - i \theta_u \right)  - i \left( \eta_u \Omega^2 + \delta \Omega \right) - 2i(|u|^2+|v|^2) &  -2i |u| |v| & -2i |u| |v| \\
\end{matrix}
\\
%\rule{11cm}{0.1pt}
%\\
\begin{matrix}
2i |u| |v| & 2i |u| |v| & -\lambda_t - \left(1 + i \theta_v \right) + i \left( \eta_v \Omega^2 + \delta \Omega \right) + 2i(|u|^2+|v|^2) & i |v|^2 \\ 
-2i |u| |v| & -2i |u| |v| & -i |v|^2 & -\lambda_t - \left(1 - i \theta_v \right) - i \left( \eta_v \Omega^2 - \delta \Omega \right) - 2i(|u|^2+|v|^2)
\end{matrix}
\end{aligned} \right|  = 0,
\label{eq:5}
\end{equation}
\end{widetext}

% \begin{equation}
% \begin{vmatrix}
% (1) & (5) & (7) & (7) \\
% -(5) & (2) & -(7) & -(7) \\
% (7 & (7) & (3) & (6) \\
% -(7) & -(7) & -(6) & (4)
% \end{vmatrix}
% \end{equation}

% \noindent with:
% \begin{enumerate}[label={(\arabic*)}]
% \item $-\lambda_t - \left(1 + i \theta_u \right) + i \left( \eta_u \Omega^2 - \delta \Omega \right)$,
% \item $-\lambda_t - \left(1 - i \theta_u \right)  - i \left( \eta_u \Omega^2 - \delta \Omega \right)$,
% \item $-\lambda_t - \left(1 + i \theta_v \right) + i \left( \eta_v \Omega^2 + \delta \Omega \right)$,
% \item $-\lambda_t - \left(1 - i \theta_v \right) - i \left( \eta_v \Omega^2 + \delta \Omega \right)$,
% \item $i |u|^2$,
% \item $i |v|^2$,
% \item $2i |u| |v|$,
% \end{enumerate}
% and

\noindent where $|u|$, $|v|$ are the fixed points of Eqs. \ref{eq:4a} and \ref{eq:4b}. The resultant fourth order polynomial in $\lambda_t$ is then solved to find roots with positive real parts (corresponding to gain) for certain normalized frequencies $\Omega$. The situation with gain and $\Omega \neq 0$ is consistent with the onset of MI, whereas $\Omega = 0$ checks for the dynamical stability to homogeneous perturbations.

\begin{figure}[!t]
\centering
\captionsetup[subfloat]{justification=centering}
\subfloat[]{\includegraphics[width=0.5\linewidth]{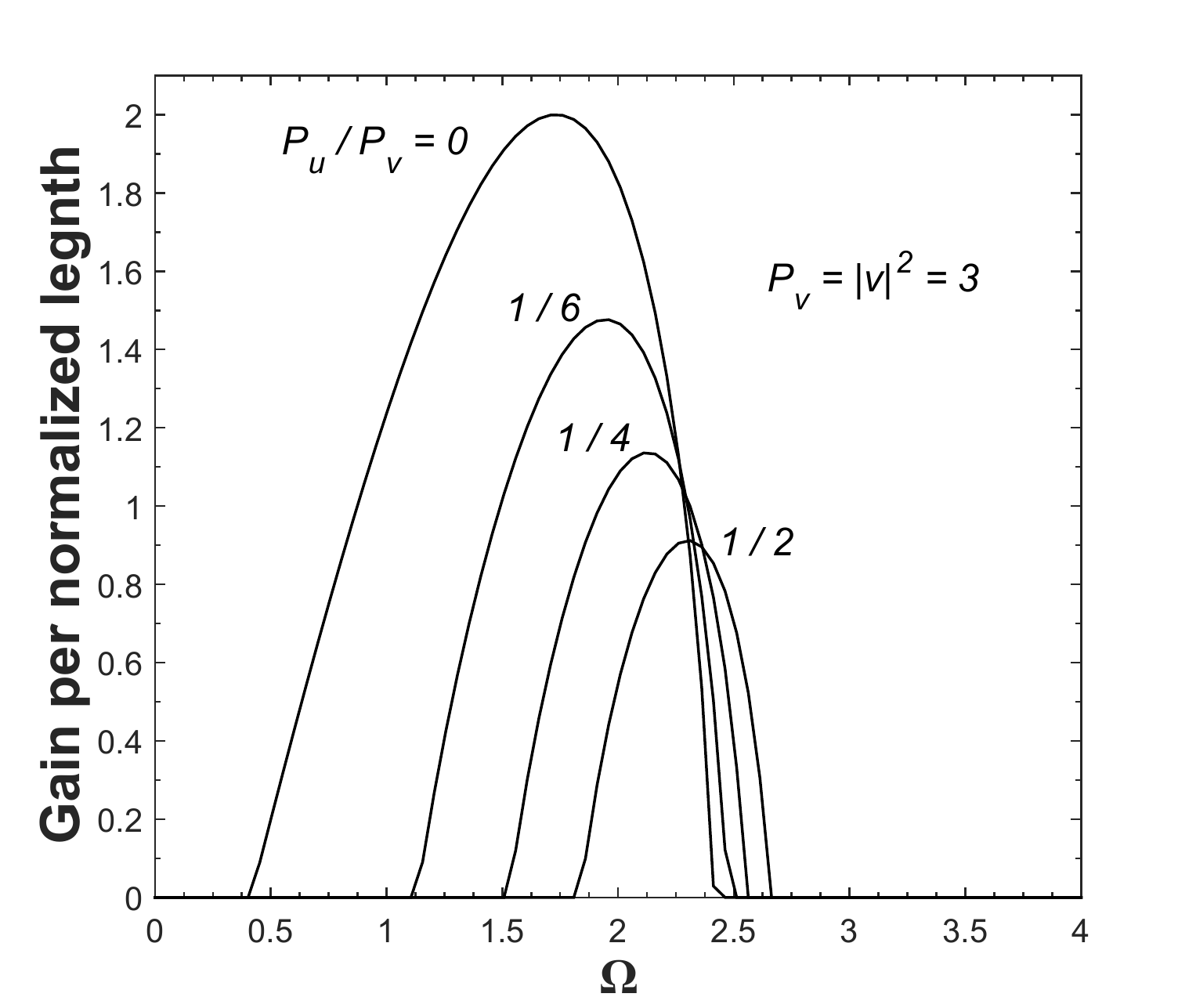}}
\subfloat[]{\includegraphics[width=0.5\linewidth]{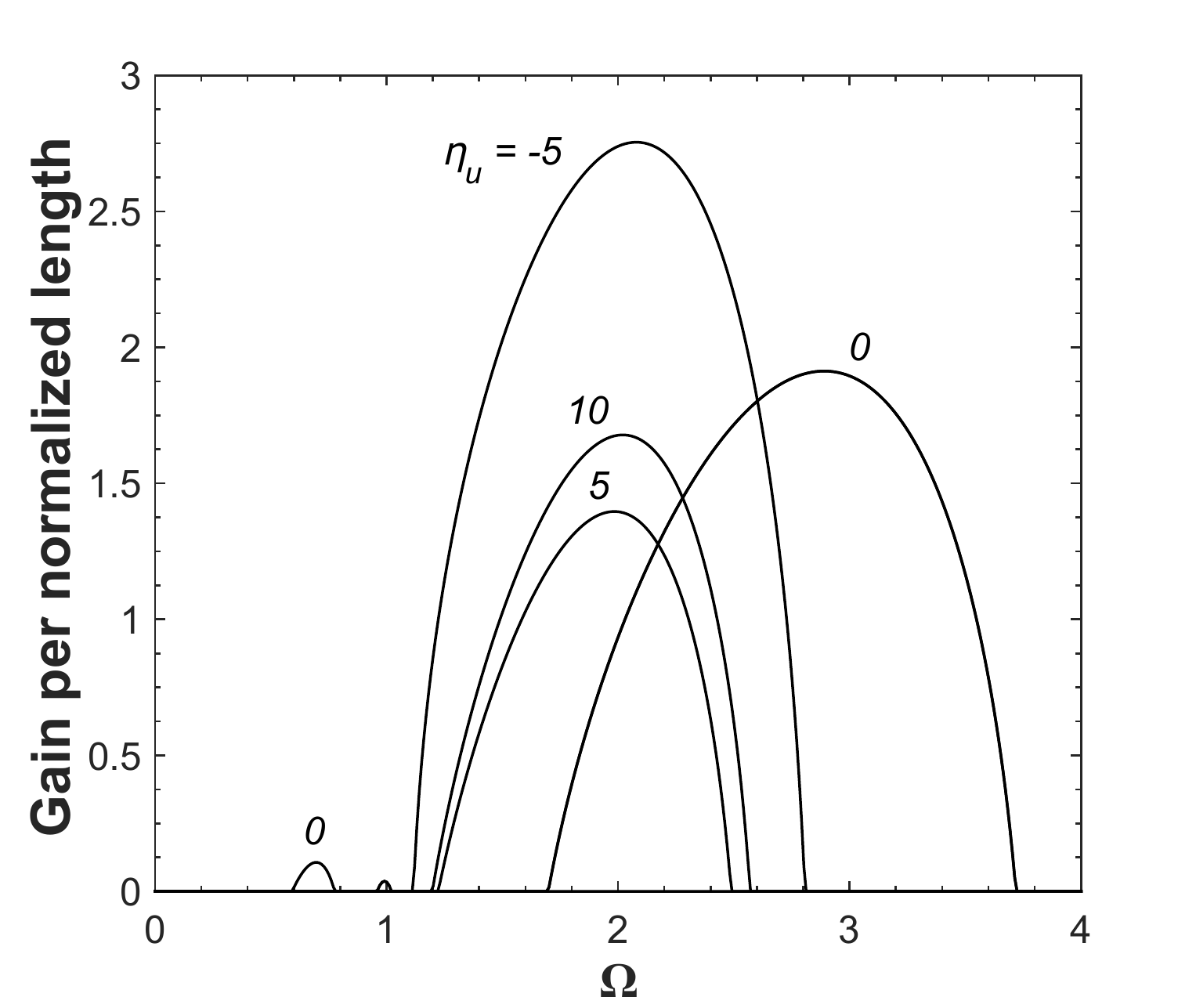}}\\
\subfloat[]{\includegraphics[width=0.5\linewidth]{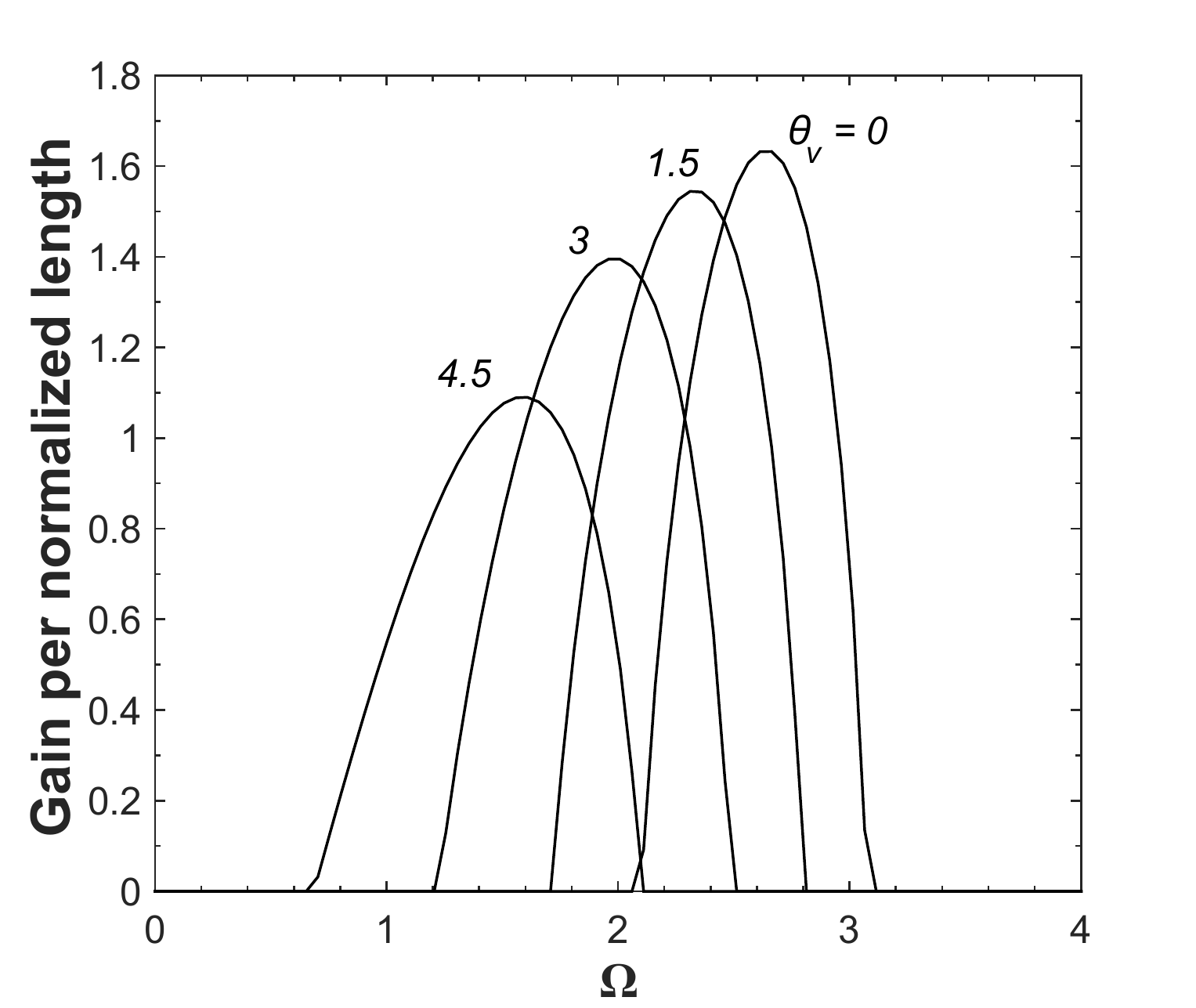}}
\subfloat[]{\includegraphics[width=0.5\linewidth]{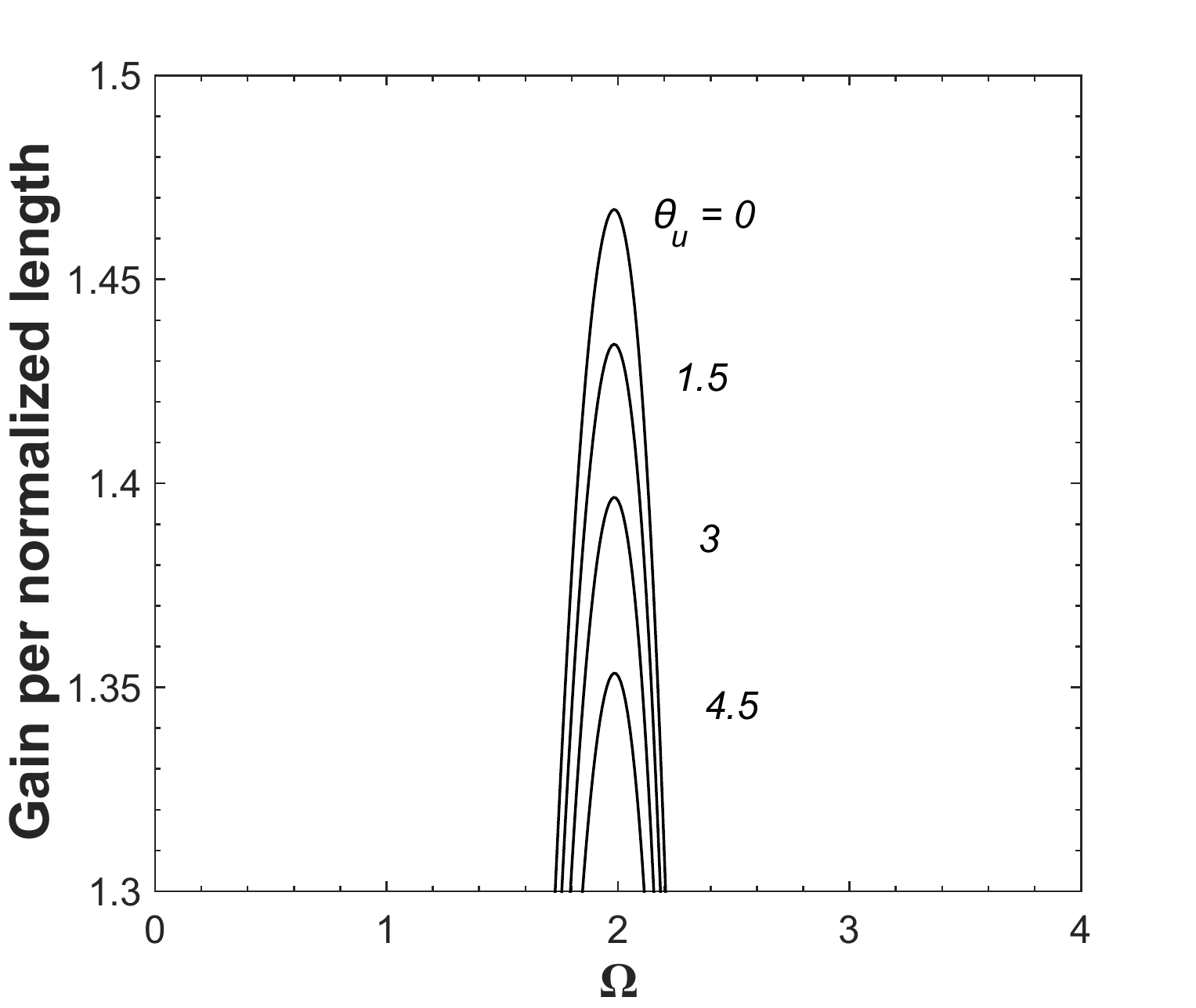}}
\caption{Impact on the MI gain of (a) intracavity power and (b) dispersion of the secondary symmetric mode $\protect{u}$; (c) detuning $\protect{\theta}$ of the primary anti-symmetric mode $\protect{v}$; and (d) detuning of the mode $\protect{u}$. Nominal parameters are $\protect{|u|^2 = 0.6}$, $\protect{|v|^2 = 3}$, $\protect{\eta_u = 5}$, $\protect{\eta_v = -1}$, $\protect{\theta_u = \theta_v = 3}$, and $\protect{\delta = 0}$.}
\label{fig:3}
\end{figure}

The full expression for these roots is too lengthy to be included in full here, so the impact of the different parameters contained within Eq. \ref{eq:5} is instead shown in Fig. \ref{fig:3}. To allow a better comparison to the conventional single-mode case, the analysis is done with a high intracavity power mode experiencing a fixed anomalous dispersion $\eta = \partial^2_\omega \beta < 0$, with $\beta$ the propagation constant and $\omega$ the optical angular frequency. We pick the anti-symmetric mode $v$ as this mode, due to $\beta = \beta_0 - \kappa_{vu} $ as a first approximation for a coupled structure, and due to $\partial^2_\omega \kappa_{vu} = \partial^2_\omega \kappa_{uv} > 0$ being typical \cite{soltani2016enabling}. The condition for MI, which nominally requires the normalized intracavity power to be unity, is only slightly perturbed, but the magnitude of the gain is generally lowered. While this indicates a reduction of the efficiency of the underlying nonlinear process, it simultaneously extends the range before the onset of spatiotemporal chaos.
 
An interesting question then becomes the fate of dissipative solitons. For this, we turn our attention to \textit{spatial} dynamics. As localized patterns are stationary with regard to the fast time variable $t$, we are able to set $\partial_t = 0$ in Eqs. \ref{eq:3a} and \ref{eq:3b} and apply perturbations of the form $\varepsilon \propto \exp\left(\lambda_s\tau\right)$ to study the possible interaction of co-existing solutions colliding with each other. This leads to

\begin{widetext}
\small
\begin{equation}
\left|
\begin{aligned}
\begin{matrix}
- \left(1 + i \theta_u \right) - i \eta_u \lambda_s^2 - 2 \delta \lambda_s + 2i(|u|^2+|v|^2) & i |u|^2 & 2i |u| |v| & 2i |u| |v| \\ 
-i |u|^2 & - \left(1 - i \theta_u \right)  + i \eta_u \lambda_s^2 - 2 \delta \lambda_s - 2i(|u|^2+|v|^2) &  -2i |u| |v| & -2i |u| |v| \\
\end{matrix}
\\
%\rule{11cm}{0.1pt}
%\\
\begin{matrix}
2i |u| |v| & 2i |u| |v| & \left(1 + i \theta_v \right) - i \eta_v \lambda_s^2 + 2i(|u|^2+|v|^2) & i |v|^2 \\ 
-2i |u| |v| & -2i |u| |v| & -i |v|^2 & - \left(1 - i \theta_v \right) + i \eta_v \lambda_s^2 - 2i(|u|^2+|v|^2)
\end{matrix}
\end{aligned} \right|  = 0.
\label{eq:6}
\end{equation}
\end{widetext}%
The roots of the resultant polynomial in $\lambda_s$, found numerically, are the eigenvalues of the spatial dynamics. The group velocity mismatch $\delta$ is kept within $u$ to ensure that the dynamics of MI for $v$ in its own moving reference frame is preserved for the case $|u| = 0$, again to allow a better comparison to the nominal single-mode case. For purely imaginary eigenvalues, the HSS oscillates and is unable to lock itself to a patterned state: this is precisely the case for the subcritical pattern in the monostable regime of the single-mode Kerr resonator. For purely real eigenvalues, single or closely packed solitons with monotonic tails are expected to form. For complex eigenvalues, soliton trains are able to emerge due to the locking of their oscillatory tails that avoid their merging. Comprehensive discussions on the nature and significance of these eigenvalues can be found in \cite{colet2014formation,parra2014dynamics}.

\begin{figure*}[!htb]
\centering
\captionsetup[subfloat]{justification=centering}
\subfloat[]{\includegraphics[width=0.33\linewidth]{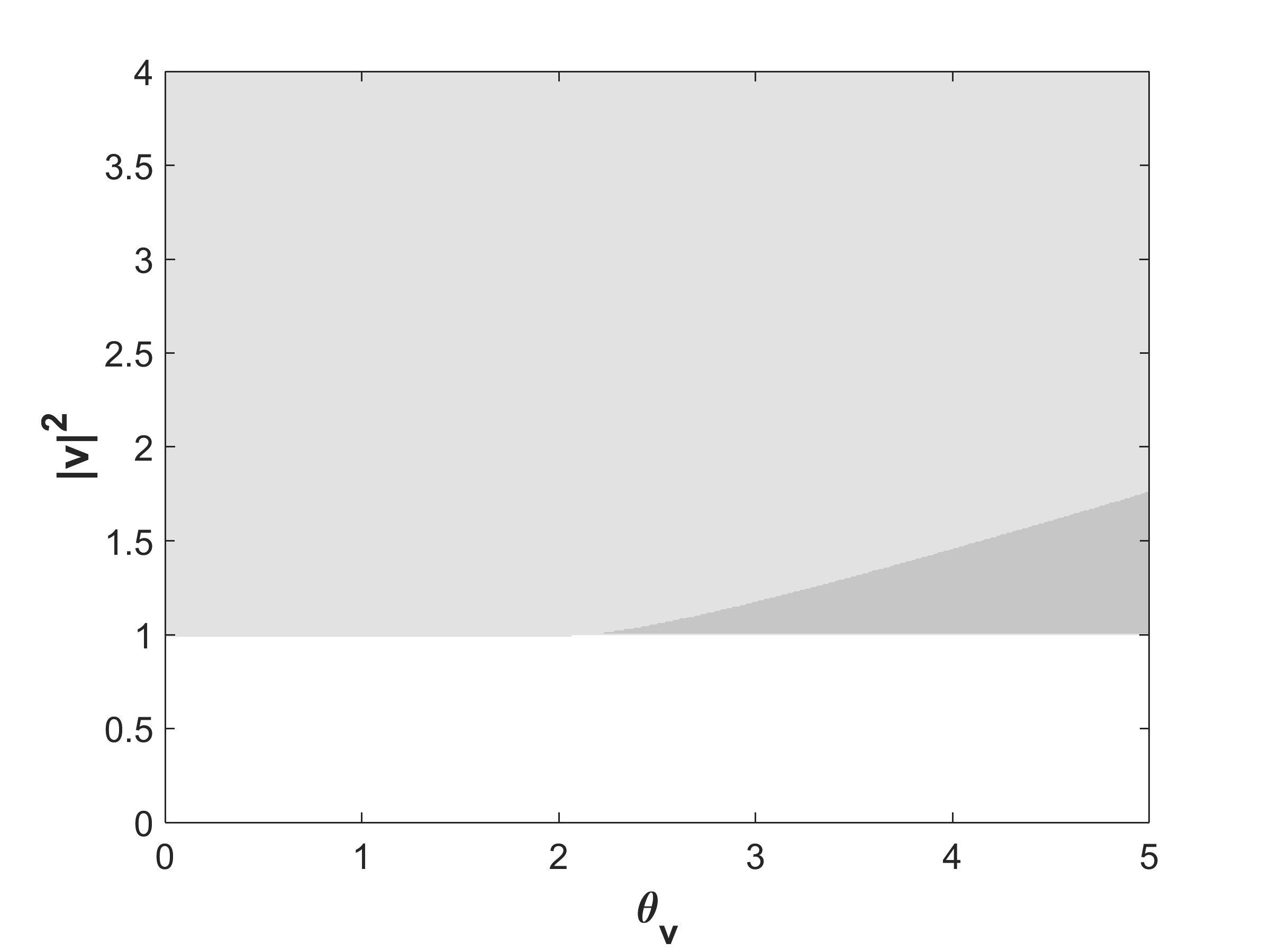}}
\subfloat[]{\includegraphics[width=0.33\linewidth]{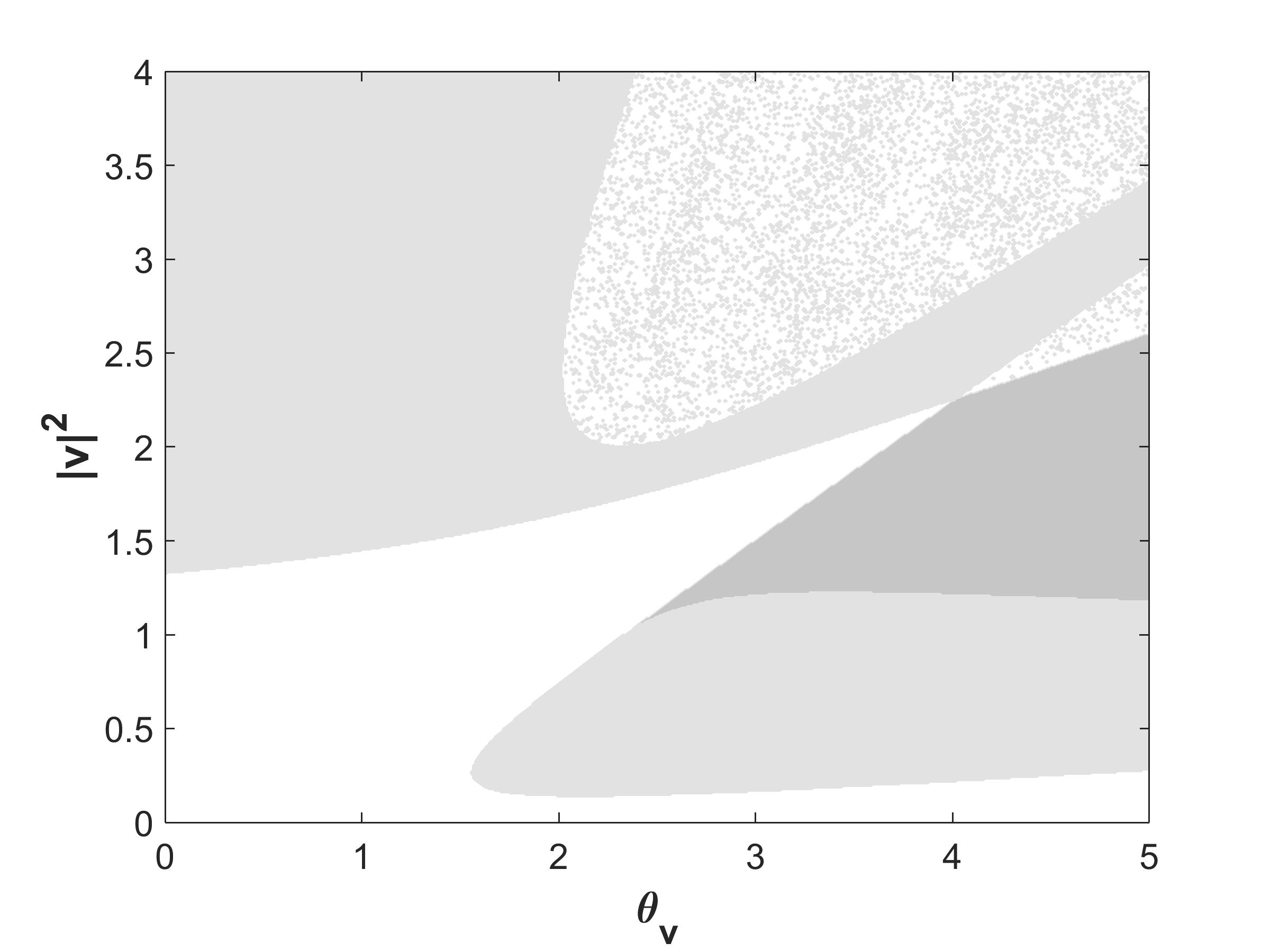}}
\subfloat[]{\includegraphics[width=0.33\linewidth]{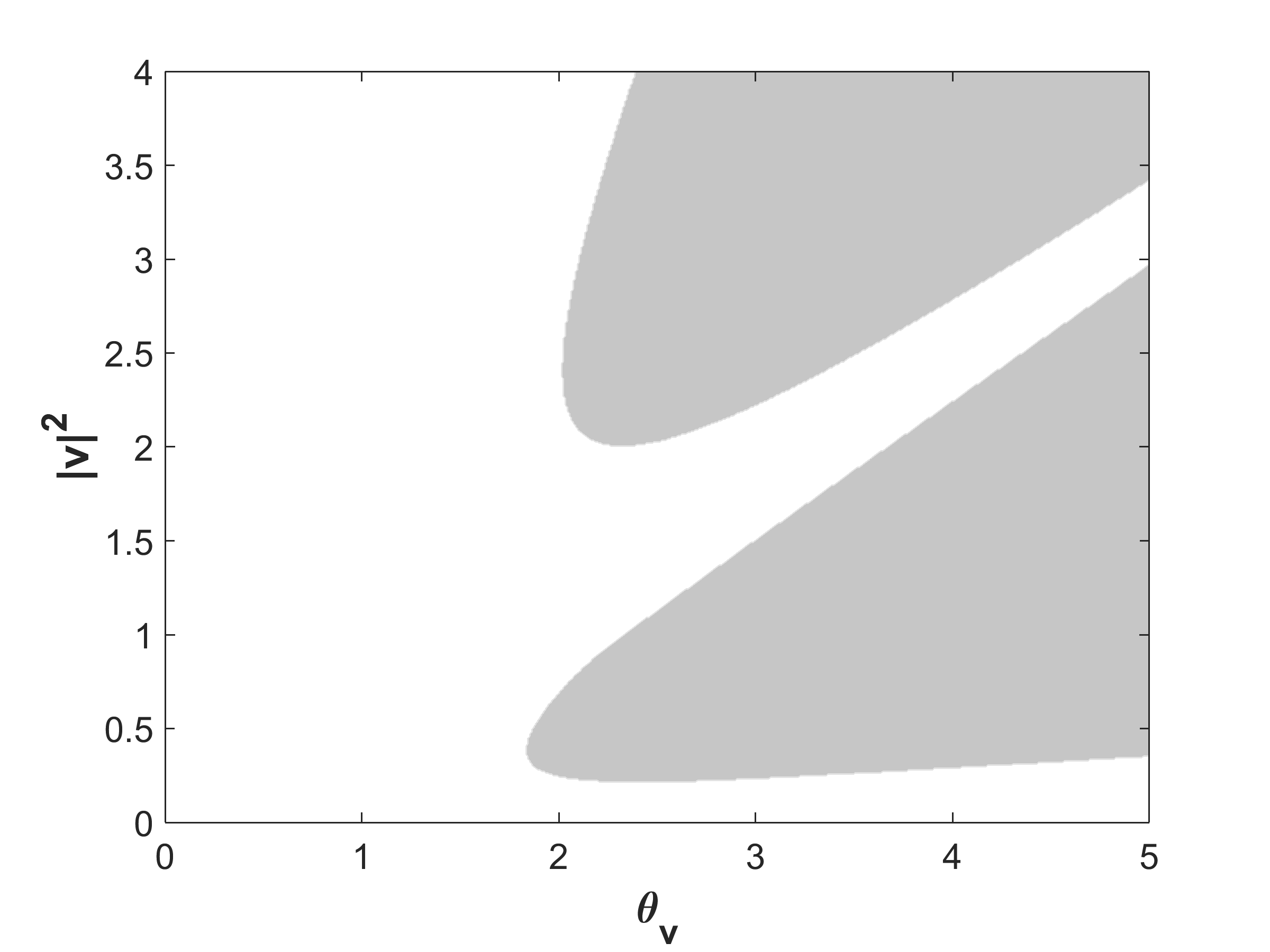}}
\caption{Regions of the parameter space ($\protect{\theta_v, |v|^2}$) with qualitatively different eigenspectra for (a) $\protect{|u|^2 = 0}$, $\protect{\delta = 0}$; (b) $\protect{|u|^2 = 0.6}$, $\protect{\delta = 0}$; and (c) $\protect{|u|^2 =  0.6}$, $\protect{\delta = 4 \times 10^{-22} \simeq 0}$. Other parameters are fixed as $\protect{\eta_u = 5}$, $\protect{\eta_v = -1}$, $\protect{\theta_u = 3}$. The white region corresponds to all the eigenvalues being complex, the light gray region to a subset being purely imaginary, and the dark gray region to a  subset being purely real. The upper right quadrant of (b) shows fast transitions between the white and light gray regions.}
\label{fig:4}
\end{figure*}

In Fig. \ref{fig:4}, similar to Ref. \cite{parra2014dynamics}, we show regions of the parameter space with qualitatively different eigenspectra -- first for the nominal case $|v| = 0$, and then for the progressive introduction of a second coupled mode with finite power and a group velocity mistmatch $\delta$. For very modest values of both, the parameter space is shown to be fully stabilized.

Whether a stable periodic pattern or a HSS is sustained in the first place has been absent from our analysis. This is instead shown with Eq. \ref{eq:5} and Fig. \ref{fig:3}. In practice, spatio-temporal chaos can still be reached for relatively modest intracavity powers, depending on the full set of parameters of Eq. \ref{eq:3}. Exploring the density of co-existing HSS with periodic patterns is complicated by the non-analytical solutions to and large dimensionality of Eq. \ref{eq:4}. More rigorously, the onset of MI should be investigated for particular physical implementations to identify sub- or super-critical behaviors and their trajectories in phase space. The linearization of the differential equations here presents a limited picture of the spatiotemporal dynamics, which are expected to be complex due to multiple bistabilities that can be triggered by large amplitude extents of the localized patterns. 

\begin{figure}[!h]
\centering
\captionsetup[subfloat]{justification=centering}
\subfloat[]{\includegraphics[width=0.5\linewidth]{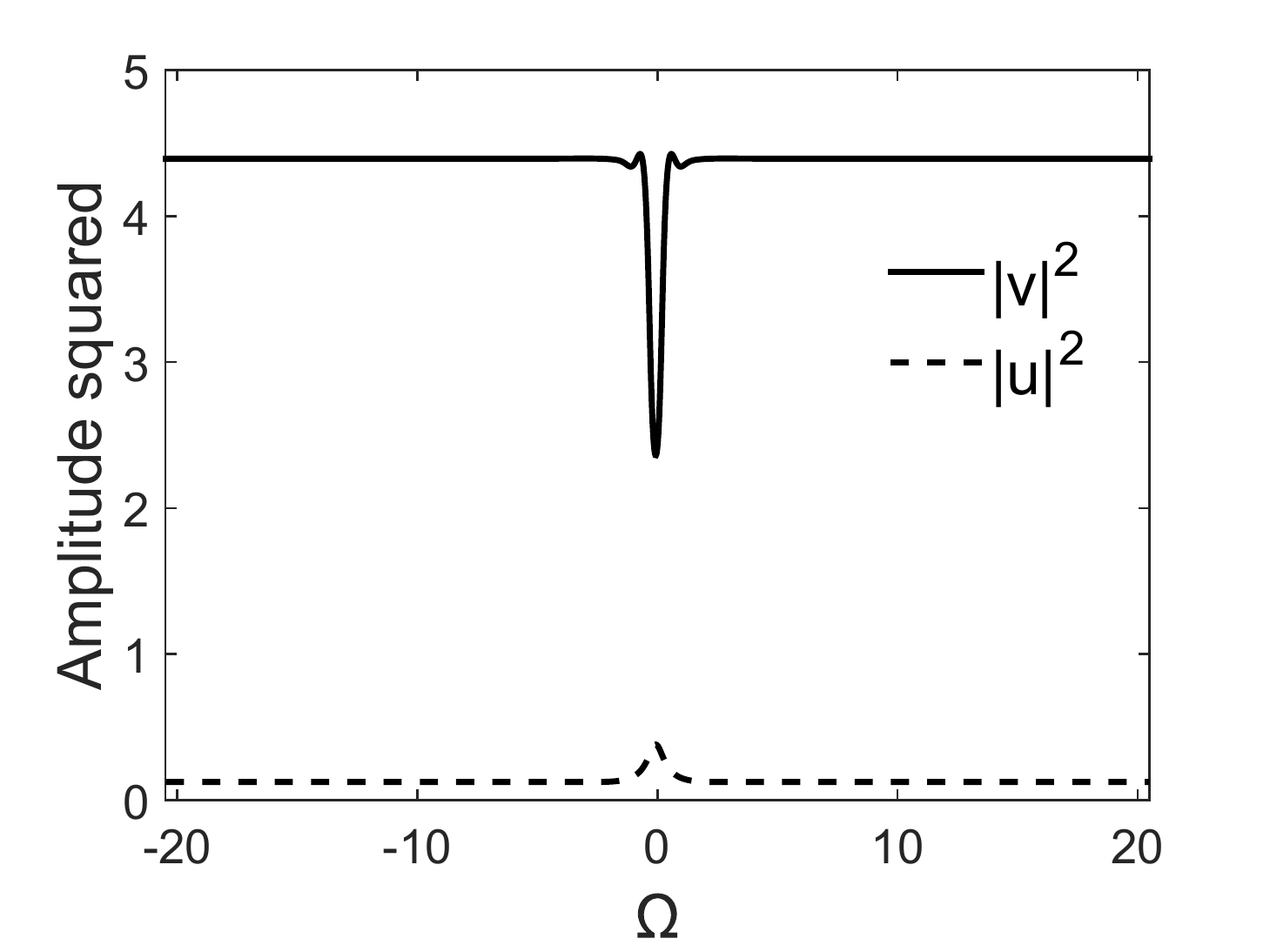}}
\subfloat[]{\includegraphics[width=0.5\linewidth]{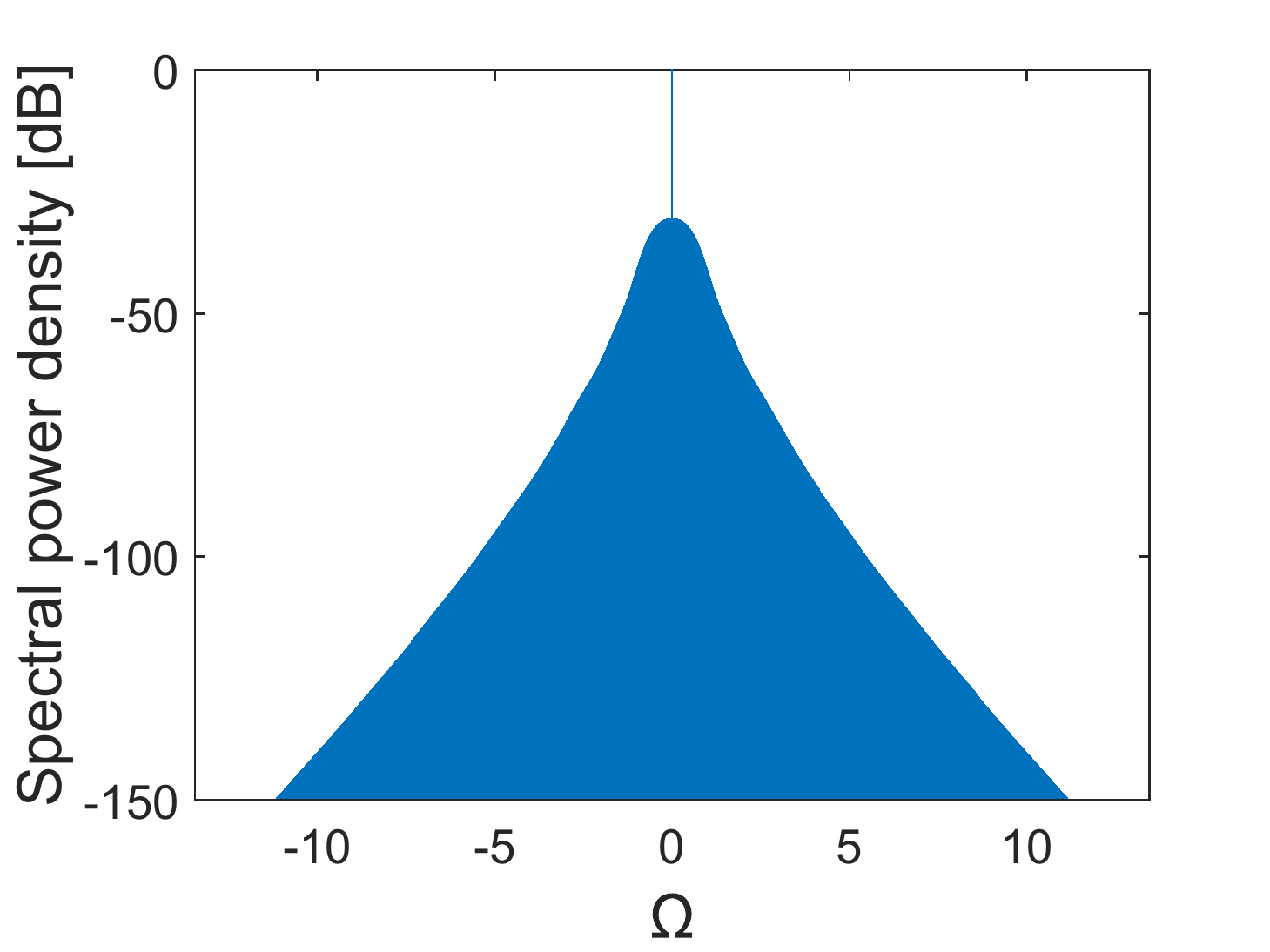}}\\
%\subfloat[]{\includegraphics[clip, trim=3cm 9cm 4cm 9cm, width=0.5\linewidth]{fig5c}}
%\subfloat[]{\includegraphics[clip, trim=3cm 9cm 4cm 9cm, width=0.5\linewidth]{fig5d}}
\subfloat[]{\includegraphics[width=0.5\linewidth]{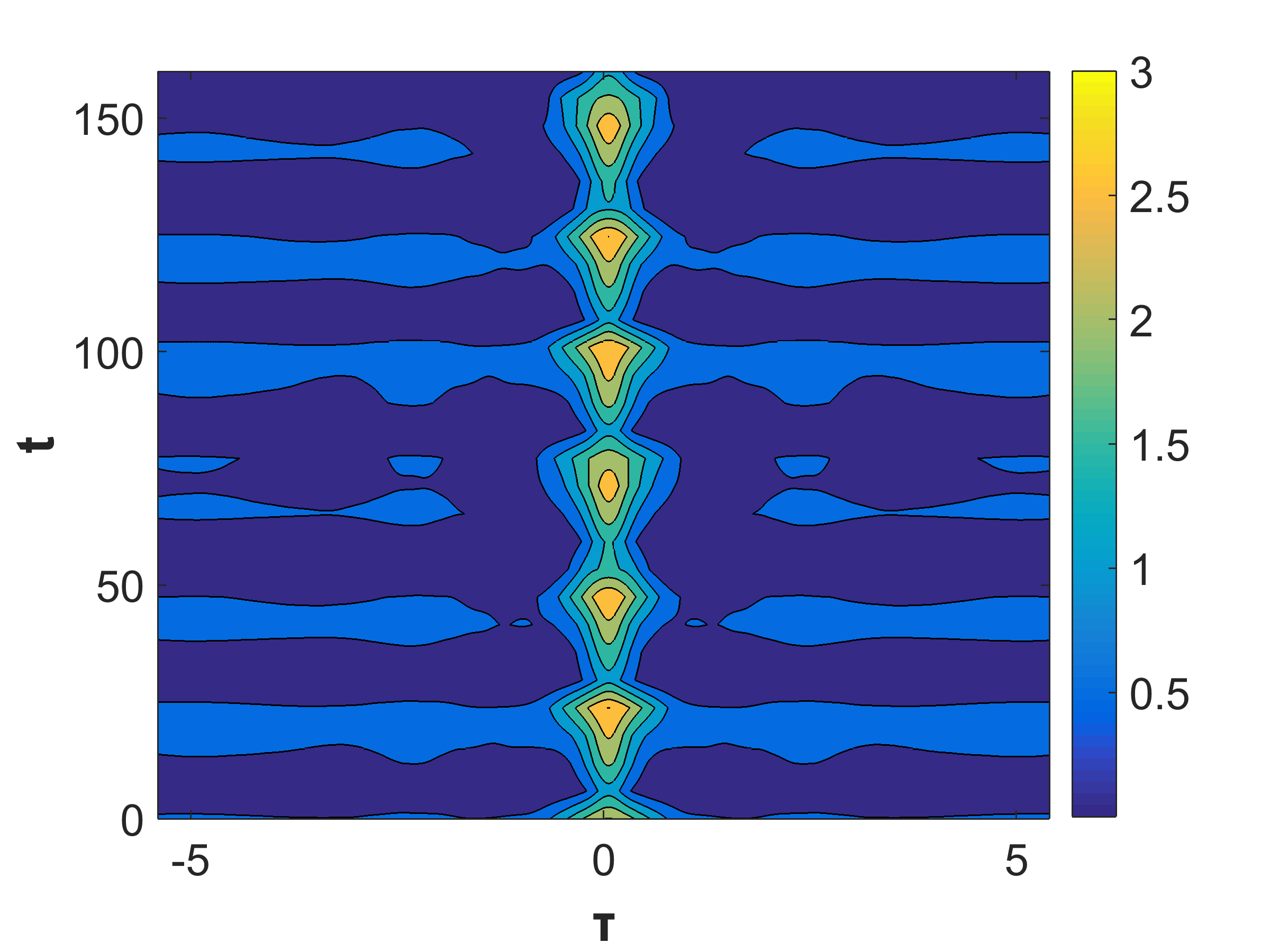}}
\subfloat[]{\includegraphics[width=0.5\linewidth]{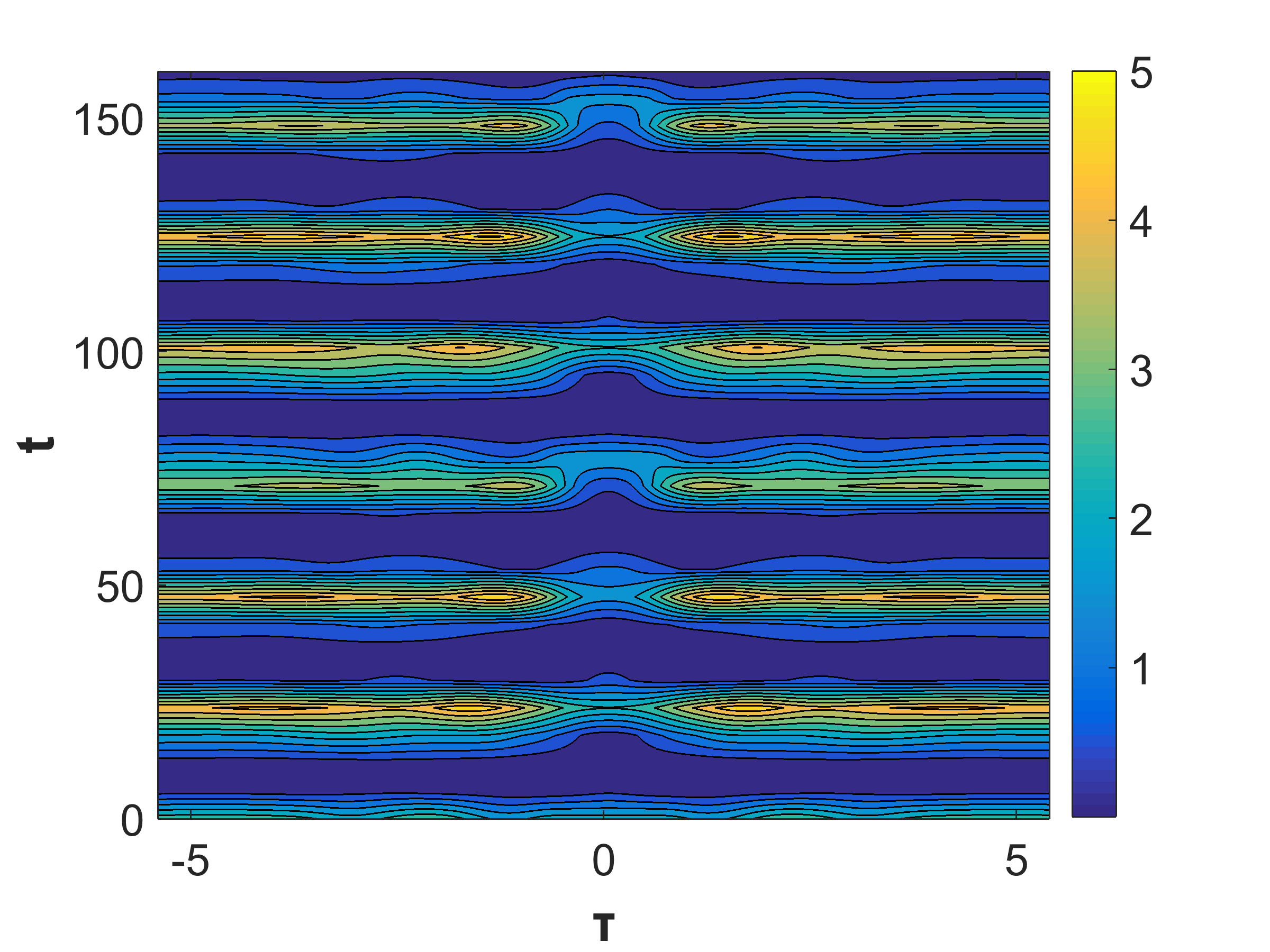}}
\caption{(a) Temporally stable dark/bright soliton pair showing $\protect{|u|^2}$ and $\protect{|v|^2}$, and (b) amplitude spectrum of the high intracavity power component $\protect{v}$, for $\protect{\eta_u = 1.84}$, $\protect{\eta_v = -0.16}$, $\protect{\theta_u = \theta_v = 4}$, $\protect{S_u = 1.76}$, $\protect{S_v = 2.49}$, and $\protect{\delta = 4.37 \times 10^{-22}}$; (c)--(d) temporal dynamics of a breather dark/bright soliton pair, showing $\protect{|u|^2}$ and $\protect{|v|^2}$ respectively, for $\protect{\eta_u = 1.84}$, $\protect{\eta_v = -0.16}$, $\protect{\theta_u = \theta_v = 3}$, $\protect{S_u = 1.82}$, $\protect{S_v = 2.03}$, and $\protect{\delta = 4.37 \times 10^{-22}}$.}
\label{fig:5}
\end{figure}

Nonetheless, by direct simulation of Eq. \ref{eq:3} with a split-step Fourier method \cite{agrawal2007nonlinear}, we find that soliton formation occurs for a large set of parameters that are robust to second-order dispersion. Fig. \ref{fig:5} shows examples of such solutions in normalized units. In terms of the sensitivity of soliton pairing to minute time delays, through a perturbation of the solutions, we observe a return to a paired state analogous to soliton trapping in birefringent optical fibers \cite{islam1989soliton}, indicating the soliton pairing is stable.

In conclusion, we have shown the formation dynamics of periodic patterns and solitons in a double-layer Kerr resonator with two coupled modes interacting via XPM are modified compared to those from a single mode. The new degrees of freedom are promising for the practical generation of Kerr frequency combs from cavity solitons, achieved through a stabilization of spatial dynamics. Future investigations can include the effects caused by higher order dispersion and the analytical studies of MI and other bifurcations in coupled structures.

% \section*{Funding}
% The financial support from the Natural Sciences and Engineering Research Council of Canada (NSERC) and Canada Research Chairs program is gratefully acknowledged.

% Bibliography
\bibliography{sample}

\end{document}